\def\year{2016}\relax
\documentclass[letterpaper]{article}
\usepackage{aaai16}
\usepackage{times}
\usepackage{helvet}
\usepackage{courier}

\usepackage{graphics}
\usepackage{algorithm}
\usepackage{algorithmic}
\usepackage{amsmath}
\usepackage{amsfonts}

\newcommand{\eat}[1]{}

\newtheorem{dfn}{Definition}

\begin{document}
% The file aaai.sty is the style file for AAAI Press 
% proceedings, working notes, and technical reports.
%

\title{Group Rotation Type Crowdsourcing}

\author{Katsumi Kumai\\
        University of Tsukuba\\
        katsumi.kumai.2015b@mlab.info
 \And Yuhki Shiraishi\\
Tsukuba University of Technology\\
 yuhkis@a.tsukuba-tech.ac.jp
 \And Jianwei Zhang\\
Tsukuba University of Technology\\
zhangjw@a.tsukuba-tech.ac.jp
 \AND Hiroyuki Kitagawa\\
University of Tsukuba\\
kitagawa@cs.tsukuba.ac.jp
\And Atsuyuki Morishima\\
University of Tsukuba\\
mori@slis.tsukuba.ac.jp
}

\nocopyright

\maketitle
\begin{abstract}
A common workflow to perform a continuous human task stream is to divide workers into groups, have one group perform the newly-arrived task, and rotate the groups. 
\eat{Usually, more than one worker belongs to each group for improving the quality of task results.}
We call this type of workflow the {\it group rotation}.
This paper addresses the problem of how to manage {\it Group Rotation Type Crowdsourcing}, the group rotation in a crowdsourcing setting.
In the group-rotation type crowdsourcing,  we must change the group structure dynamically because workers come in and leave frequently. %However, changing the group structure will give workers psychological stress\eat{, such as surprise, confusion or irritation}. 
This paper proposes an approach to explore a design space of methods for group restructuring in the group rotation type crowdsourcing.
% and compares the methods in terms of the evaluation results on psychological stress with real-world crowd workers.

\end{abstract}

\section{Introduction}\label{intro}

Continuous human task streams appear in many applications, such as the captioning of real-time broadcasting and the metadata labeling to objects in videos \cite{LMS+12} \cite{NGL+13}.
An example of task in such a task stream is to transcribe one spoken sentence into text.

Since human resources are limited, a common workflow to perform such a task stream is to divide workers into groups, have one group perform the newly-arrived task, and rotate the groups \cite{WFD}. 
%{\bf Some professional interpreters often use this workflow to reduce the mistake by mental or physical fatigue \cite{WFD}.}  
\eat{Workers in groups that do not perform tasks can take a rest until their turn comes.}
In general, more than one worker belongs to each group for improving the result quality.
We call this type of workflow the {\it group rotation}.

\begin{figure}[t]
\begin{center}
\scalebox{0.18}{
  \includegraphics{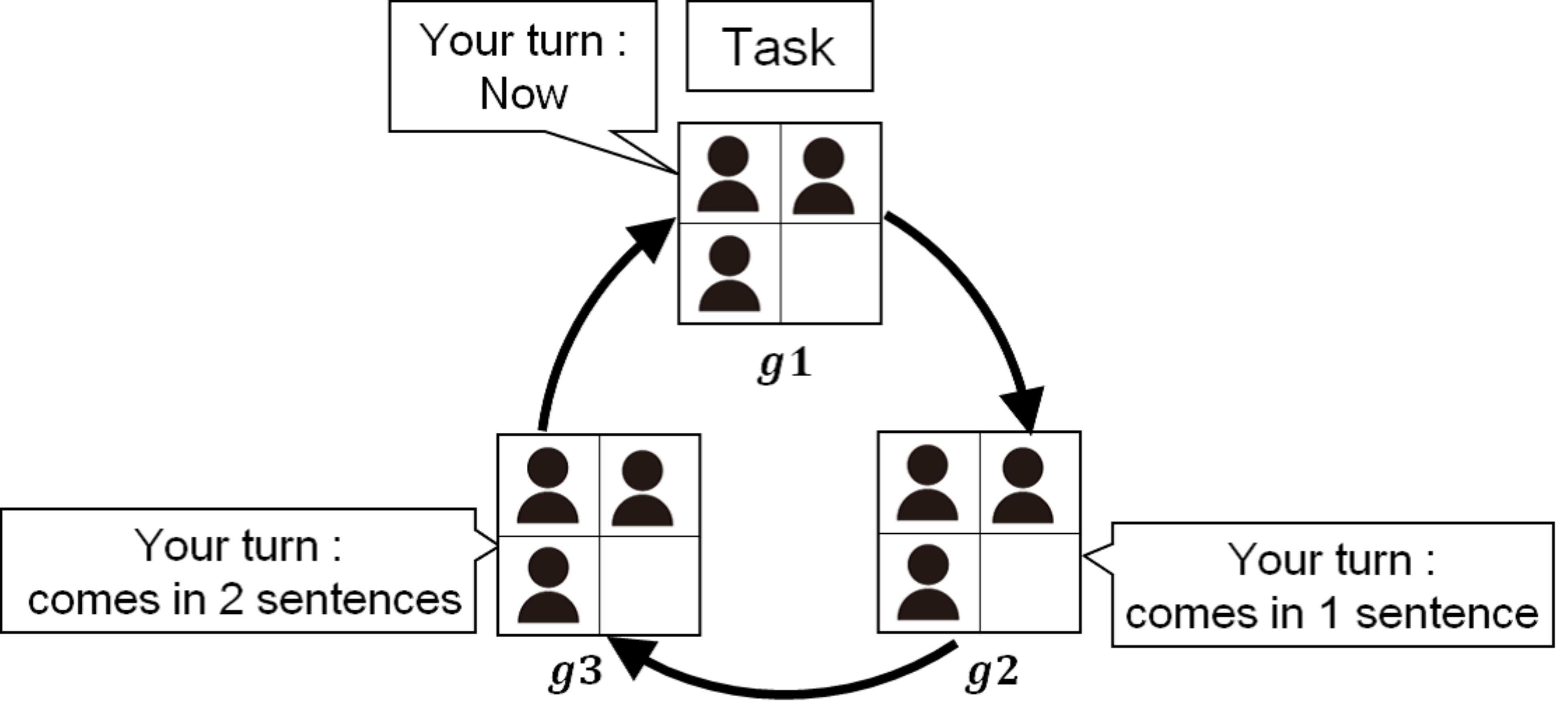}
}
\vspace{-2mm}
  \caption{Group rotation ($d=2$)}
  \label{fig:fig1}
  \end{center}
\end{figure}

Fig. \ref{fig:fig1} illustrates a group rotation.
Assume that we have a task stream for transcribing sentences spoken in a video.
We have three groups $g_1$, $g_2$ and $g_3$. At present, workers in $g_1$ are performing the task.
Each task asks workers to transcribe one sentence. Their results in a group will be aggregated for improving the task result by some means (e.g., majority voting).
Then, workers in $g_2$ will transcribe the next sentence.

There are two points here.
First, it is important to let workers know when their turn comes.
Therefore, we put a {\it counter} on the task screen of each worker that countdowns until the sentence the worker has to transcribe appears. 
With the counter, workers can prepare for their turn.

Second, there is an application-specific number $d\geq 1$, which is the minimum number of workers in each group.
In Fig. \ref{fig:fig1}, $d=2$. 
Usually, the minimum number of workers in a group is determined by the way how the application aggregates the answers 
to maintain the quality of task results.  

%$d$ must beThe number of workers in a group, is not less than a specific number $d$. ($d$ depends on the quality of the task results that the task requester wants to guarantee).

%図は小さくする

{\bf Group Rotation Type Crowdsourcing.}
This paper addresses the problem of how to manage {\it Group Rotation Type Crowdsourcing (GRTC)}, the group rotation in a crowdsourcing setting.
Our assumption is that we can always recruit workers during the task stream, allowing workers to come in  and leave freely.
An example is to recruit volunteer workers from the audience of a lecture for transcribing the lecture.

Under the assumption, we must change the group structure dynamically because workers come in and leave while tasks are being performed.
While it is desirable to increase the number of groups to make the burden to workers small, we must reduce the number of groups when there is a group having less than $d$ workers.

However, changing the group structure will give workers psychological stress, such as surprise, confusion or irritation. 
For example, if the counter jumps from 20 to 2, the worker would be surprised and feel stressed since she may not have prepared for the task.
There is a clear tradeoff between optimizing the number of groups dynamically and keeping the psychological stress of workers small. 
The problem may look similar to those for tree-form database indices such as B-Trees \cite{Comer79} \cite{BayerM70}.
In such index structures, we usually address the tradeoff between the access time and the required space for storing the index.
In contrast, a unique point of our problem is that the target is humans and not data. 
%They would feel stressed if  they were moved in the group rotation in unexpected ways.
We address the tradeoff between optimizing the number of groups and keeping the psychological stress given to workers small.

This paper proposes an approach to explore a design space of methods for group restructuring in GRTC.
Our purposes are to (1) confirm the tradeoff between increasing the number of groups and keeping the psychological stress of workers caused by the move to other groups small, and (2) hopefully find sweet spots in the tradeoff.
% and compares the methods in terms of an evaluation result on psychological stress with real-world crowd workers.}

%{\bf Our contribution is modeling the group rotation in Crowdsourcing. To the best of our knowledge, there is no work that explicitly addresses the problems that happen if we apply group rotations to crowdsourcing settings. This paper explores a design space of methods for group restructuring in the group rotation type crowdsourcing}

%ここにcontribution追加

%\section{Group Rotations}\label{grtc}
%This section gives definitions used to discuss group rotations.

\section{Group Rotations}
A {\it group rotation state} (or shortly a {\it grs}) is a building block of a group rotation and represents a snapshot of it.
Fig. \ref{fig:fig1} illustrates a grs.
Formally, a grs is defined as follows:

\begin{dfn}
A group rotation state $S$ is defined as a tuple $(W, G, W_g, Succ, p)$ where:

\begin{itemize}
\item $W = \{w_1,w_2,\cdots,w_n\}$ is a set of workers. $|W|=9$ in Fig. \ref{fig:fig1},
\item $G = \{g_1,g_2,\cdots,g_m\}$ is a set of groups ($|G| \geq 2$). $G=\{g_1, g_2, g_3\}$ in Fig. \ref{fig:fig1},
\item $W_g: G \to 2^W$ maps each group to the workers who belongs to it. For example, $|W_g(g_1)|=3$ in Fig. \ref{fig:fig1},
\item $Succ: G \to G$ defines the next group of each group for the rotation. $Succ(g_1)=g_2$ in Fig. \ref{fig:fig1},
The function is illustrated by direct edges among nodes representing groups, and 
\item $p \in G$ is the group whose workers are performing a task at this state. $p=g_1$ in Fig. \ref{fig:fig1}.\hfill$\Box$
\end{itemize}
%We also use $W_{g_x}$  to denote a set of workers who belongs to $g_x$ (i.e., $W_{g_x} = \{w | Belongs(w) = g_x\}$).
\end{dfn}

In $S$, (1) every worker must belong to exactly one group, (2) each group has to must have at least one worker, and (3) the graph must have a circle shape. Namely, $S$ must satisfy all the following conditions:
\begin{enumerate}
\item $W_g(g_1)\oplus W_g(g_2)\oplus \cdots \oplus W_g(g_m) = W$, 
\item $\forall g_x\in G\;(|W_g(g_x)| \geq 1)$, and
\item For any two groups $g_i, g_j\in G$, there is one and the only path from $g_i$ to $g_j$ with $Succ$. 
\end{enumerate}

Given two grs's $S$ and $S'$, we say that $S'$ {\it follows} $S$ if any worker in $p$ who performed tasks in $S$ does not perform any task in $S'$ and $p'$ is the successor of $p$ in $S'$. Formally,

\begin{dfn}
Let $S$ and $S'$ be group rotation states, and let $p$ and $p'$ be the current groups of $S$ and $S'$, respectively.
We say $S'$ {\it follows} $S$ if (1) $W_g(p)\cap W_g'(p')=\phi$, (2) $p$ exists in $G'$, and (3) $p'=Succ'(p)$.\hfill$\Box$
\end{dfn}

%Now we define a group rotation.
A group rotation is a sequence of group rotation states each of which follows its predecessor.

\begin{dfn}
Let $[S_1,S_2,\ldots]$ be a sequence of group rotation states.
The sequence is a {\it group rotation} if for any successive pair $(S_i, S_{i+1})$ in the sequence, $S_{i+1}$ follows $S_i$.\hfill$\Box$
\end{dfn}

\section{Group Rotation Generators}

Assume that  we have an application-dependent minimum number of workers for each group (denoted by $d$), a sequence $T=[t_1,t_2,\ldots,]$ of times when each task is performed, and a sequence $\Delta W = [\Delta w_{i_1}, \Delta w_{i_2}, \ldots,]$ of worker changes.
Here, $\Delta w_i$ is either $+w_i$ (i.e., $w_i$ comes in) or $-w_i$ ($w_i$ leaves) and has a property $t(\Delta w_i)$ to represent the time when the change happens.
Then, we can generate a group rotation with a {\it group rotation generator}, an algorithm to generate group rotations.

A group rotation generator is defined 
as follows:

\begin{dfn} 
The group rotation generator is defined as a function $Next:States\times Int\times Diff\rightarrow States$ that takes as input $S_i$, $d$ and a subsequence of $\Delta W$ and generates the next $S_{i+1}$ s.t. $S_{i+1}$ follows $S_i$. \hfill$\Box$
\end{dfn}

Given  $(S_1, d, T, \Delta W, Next)$, the following procedure generates a group rotation.
\begin{enumerate}
\item Output $S_1$ as the first grs in the group rotation.
\item At each $t_i \mbox{ in } T$ do the following.
\begin{enumerate}
\item Let $\Delta W_i$ be the subsequence of $\Delta W$ in which $t(\Delta w_j)$ for each $\Delta w_j\in \Delta W_i$ is in $(t_{i-1}, t_i]$. Namely, $\Delta W_i$ is a set of worker changes from $t_{i-1}$  to $t_i$.
\item $S_{i+1}=Next(S_i, d, \Delta W_i)$ where $p_{i+1}=Succ_{i+1}(p_i)$.
\end{enumerate}
\end{enumerate}

%section{Design Space for Group Rotation Generators}\label{design}

Algorithm \ref{alg:template} shows a design space for $Next(S_i, d, \Delta W_i)$.
In the design space,  we apply two worker-at-a-time update operators (named Insert and Remove) for generating $S_{i+1}$ in a sequential way according to worker insertion and deletion described in the sequence $\Delta W_i$.
The algorithm works as follows. 
First, it copies $S_i$ to $S_{tmp}$ (Line 1). Next, it applies Insert and Remove operators with each worker $\Delta w_j$ in $\Delta W_i$ (Lines 2 to 8). Finally, it copies $S_{tmp}$ to $S_{i+1}$ and moves the current group forward (Line 9).
The two operators work as follows. First, $Insert(S, d, +w_j)$ {\bf chooses} a group and inserts $w_j$ into it. If the number of workers in the group is larger than a function of $d$ (denoted by $max(d)$), it {\bf splits} the group into two groups.
Second, $Remove(S, d, -w_j)$  first deletes $w_j$ from a group. If the number of workers in the group becomes less than $d$, 
it moves other workers to the group if we {\bf find} a group having many workers, otherwise {\bf joins} the groups with another group to meet the condition.

Here, we see four key components: {\bf choose}, {\bf split}, {\bf find} and {\bf join}.
We consider a variety of possible methods to implement the four components in Insert and Remove operators.
For example, choosing a group into which we insert worker heavily affects how often the groups are restructured and how much stress workers experience.

\eat{
arent when workers are inserted to groups, since if we increase the number of groups, some workers have to move.
On the other hand, when workers are leaving, preventing groups from being joined is generally a good strategy 
since it prevents from both decreasing the number of groups and moving many workers.
}

%Exploring wider design spaces, such as a design space with holistic worker-set-at-a-time operators for the update, is our interesting future work.
%Although it is possible to explore a wider design space with holistic worker-set-at-a-time operators for the update, the design space with worker-at-a-time operators is large enough.
%Exploring wider design spaces is our interesting future work.

%the following algorithm templat.

\begin{algorithm}[t]
	\caption{Template for $Next(S_i, d, \Delta W_i)$}
	\label{alg:template}
	\footnotesize
	\begin{algorithmic}[1]
	\REQUIRE $S_i, d, \Delta W_i$
	\ENSURE $S_{i+1}$
		\STATE  $S_{tmp}\leftarrow S_i$
		\FOR{ $\Delta w_j\in \Delta W_i$}
                   \IF{$\Delta w_j$ is $+w_j$}
                     \STATE $S_{tmp}\leftarrow Insert(S_{tmp}, d, +w_j)$
                   \ELSIF{$\Delta w_j$ is $-w_j$}
                      \STATE  $S_{tmp}\leftarrow Remove(S_{tmp}, d, -w_j)$
                   \ENDIF
                \ENDFOR
		\STATE $S_{i+1}\leftarrow S_{tmp}$ with $p_{i+1}=Succ_{i+1}(p_i)$.
	\end{algorithmic}
\end{algorithm}

\section*{Acknowledgments}
The authors are grateful to the contributors to Crowd4U, whose names are partially listed at http://crowd4u.org.
This work was partially supported by JSPS KAKENHI (\#25240012,
\#26870090, \#16K16460), Collaborative Research Program at NII, Expense for
Strengthening Functions in NTUT's Budgetary Request for Fiscal 2016 and
Promotional Projects for Advanced Education and Research in NTUT.

\eat{This research was partially supported by the Grant-in-Aid for Scientific Research (\#25240012, \#26870090 and \#16K16460) from MEXT, 
the collaborative research program at NII, and Japan and Promotional Projects for Advanced Education and Research in Tsukuba University of Technology.}

\bibliographystyle{aaai}
\bibliography{references}

\end{document}